\documentclass[12pt]{article}

\usepackage{amsmath,amsfonts}
\usepackage{bm}
\usepackage{natbib,theorem}
\usepackage[dvipdf]{graphicx}
\usepackage{amssymb}
\usepackage{epsfig}
\usepackage{latexsym}
\usepackage{multicol}
\usepackage[dvips]{color}
\usepackage{setspace}
\usepackage{hyperref}

\renewcommand{\baselinestretch} {1.30}
\newtheorem{theorem}{Theorem}
\newtheorem{lemma}{Lemma}
\newtheorem{corollary}{Corollary}
\newtheorem{proposition}{Proposition}

\theoremheaderfont{\hskip\parindent\normalfont}
\theorembodyfont{\normalfont}

\newcommand{\bA}{\mathbf{A}}

\newcommand{\bB}{\mathbf{B}}
\newcommand{\bC}{\mathbf{C}}

\newcommand{\bd}{\mathbf{d}}
\newcommand{\bD}{\mathbf{D}}

\newcommand{\bF}{\mathbf{F}}
\newcommand{\bG}{\mathbf{G}}
\newcommand{\bI}{\mathbf{I}}
\newcommand{\bH}{\mathbf{H}}
\newcommand{\bh}{\mathbf{h}}

\newcommand{\bP}{\mathbf{P}}

\newcommand{\bS}{\mathbf{S}}

\newcommand{\bT}{\mathbf{T}}
\newcommand{\bU}{\mathbf{U}}
\newcommand{\bu}{\mathbf{u}}
\newcommand{\bV}{\mathbf{V}}
\newcommand{\bbv}{\mathbf{v}}
\newcommand{\bW}{\mathbf{W}}

\newcommand{\bX}{\mathbf{X}}

\newcommand{\bY}{\mathbf{Y}}

\newcommand{\bZ}{\mathbf{Z}}

\newcommand{\cN}{\mathcal{N}}
\newcommand{\cX}{\mathcal{X}}

\newcommand{\cS}{\mathcal{S}}

\newcommand{\bbeta}{\boldsymbol{\beta}}

\newcommand{\bDelta}{\boldsymbol{\Delta}}

\newcommand{\bGamma}{\boldsymbol{\Gamma}}

\newcommand{\bmu}{\boldsymbol{\mu}}

\newcommand{\brho}{\boldsymbol{\rho}}
\newcommand{\bSigma}{\boldsymbol{\Sigma}}

\newcommand{\bxi}{\boldsymbol{\xi}}

\newcommand{\rank}{\mathrm{rank}}
\newcommand{\spn}{\mathrm{span}}

\newcommand{\bzero}{\boldsymbol{0}}

\newcommand{\argmax}{\operatornamewithlimits{argmax}}
\newcommand{\argmin}{\operatornamewithlimits{argmin}}

\begin{document}

\numberwithin{equation}{section}
\renewcommand{\baselinestretch}{1.5}

\title{A New Reduced-Rank Linear Discriminant Analysis Method and Its Applications}

\author{Yue Selena Niu, Ning Hao, and Bin Dong\\
University of Arizona and Peking University}
\date{\today}
\maketitle

\begin{abstract}
We consider multi-class classification problems for high dimensional data. Following the idea of reduced-rank linear discriminant analysis (LDA), we introduce a new dimension reduction tool with a flavor of supervised principal component analysis (PCA). The proposed method is computationally efficient and can incorporate the correlation structure among the features. Besides the theoretical insights, we show that our method is a competitive classification tool by simulated and real data examples.

\end{abstract}

\section{Introduction}
Targeting on cancer classification and other modern applications, a great number of high dimensional classification techniques have been invented and studied recently; see \cite{hastie2009elements} for an extensive introduction, and \cite{witten2011penalized,cai2011direct,fan2012road,mai2012direct} for some recent developments. Although these contemporary classification tools can be applied to the high dimensional data, most of them rely on some strong assumptions. For example, many methods assume that the features are independent to each other; some other methods assume various sparsity conditions. On one hand, these assumptions make the model simple and robust against growing dimensionality, so classification accuracy and computational efficiency can be achieved. On the other hand, they may be too restrictive, and when violated, lead to information loss in data analysis. Moreover, lots of methods target on the binary classification case and are not straightforward to use if more than two classes are present. Convenient and efficient classification tools for multi-class data are quite limited. Therefore, it is desirable to develop new classification techniques that can handle high dimensional, multi-class data and also take into account the correlation among the features.

Recall that many linear classification rules depend on the Mahalanobis distance. However, it cannot be well-estimated for high dimensional data when the number of features is greater than the sample size as the sample covariance is singular. Nevertheless, under the assumption that features are independent, the sample covariance matrix is diagonal and strictly positive definite, so the Mahalanobis distance can be calculated. That is one of the main reasons that the independence assumption is crucial in many popular classification methods.  For example, the nearest shrunken centroids \citep[NSC,][]{tibshirani2002diagnosis}, independence rule \citep[IR,][]{BickelLevina:2004}, features annealed independence rule \citep[FAIR,][]{FanFan:2008} all assume that the features are independent to each other. Moreover, some other methods such as regularized discriminant analysis \citep[RDA,][]{guo2007regularized} use a covariance estimator which is the sample covariance regularized towards a diagonal matrix. Recently, many new classification tools have been developed including penalized linear discriminant analysis \citep[PLDA,][]{witten2011penalized}, linear programming discriminant rule \citep[LPD,][]{cai2011direct}, regularized optimal affine discriminant rule \citep[ROAD,][]{fan2012road}, direct sparse discriminant analysis \citep[DSDA,][]{mai2012direct}, sparse discriminant analysis \citep[SDA,][]{clemmensen2012sparse}, multi-class sparse discriminant analysis \citep[MSDA,][]{mai2015multiclass}. Roughly speaking, these sparse methods obtain sparse models by solving penalized or constrained optimization problems, and their efficiency relies on the sparsity level of the normal vectors to the optimal discriminant boundaries.

Reduced-rank LDA is a classical approach to classification. It conducts dimension reduction by projecting the data to the centroid-spanning space and classifies the data based on nearest centroid. Another commonly used dimension reduction tool is PCA, which projects the data to the space spanned by the top principal components of the total sample covariance matrix. Note that reduced-rank LDA makes use of the label information (through centroids) but ignores the (within class) covariance information. On the other hand, PCA relies on the covariance information only and is mainly regarded as an unsupervised learning tool in the literature.

In this paper, we propose a new reduced-rank LDA method combining the advantages of the classical reduced-rank LDA and PCA. The principal components of a weighted sum of the sample within class and between class covariance matrices are used for dimension reduction, and standard LDA is employed to the projected data for classification. In this dimension reduction process, both label and covariance information can be taken into account, through between class and within class covariance, respectively.  Therefore, we regard it as a version of supervised PCA. This new method does not rely on the aforementioned sparsity or independence assumptions and offers an alternative classification tool for various applications, especially, when these technical assumptions are not satisfied.

To shed theoretical insight of our method, we consider the spiked structure of the covariance \citep{johnstone2001distribution}. Roughly speaking, a symmetric positive definite matrix is called spiked if all of its eigenvalues are equal except a few large ones. In other words, it is a sum of a scalar matrix and a low rank matrix. Intuitively, the spiked structure might be a better model than the diagonal one to approximate the true covariance, as it can take into account strong correlation among the features, which is not uncommon in applications. We will explain why the spiked structure can help for dimensional reduction. We should remark that, although the spiked structure plays a key role in the theoretical analysis, our proposed procedure does not rely on the spiked structure explicitly and can be applied to general data sets.

The main contribution of this paper is threefold. First, it proposes a novel dimension reduction and classification tool. The new method incorporates covariance among features and works well for high dimensional multi-class data. Second, it illustrates a new supervised way to conduct PCA. This technique is generally applicable to both classification and regression models. Last but not least, the proposed method is computationally efficient and can be applied directly to the real data such as gene expression data in cancer research.

The rest of the paper is organized as follows. Section 2 introduces notations and gives a selective review on some popular linear classification methods from an aspect of dimension reduction. In Section 3, we study a new reduced-rank LDA method for classification and offer some theoretical insights. Numerical studies and real data applications are illustrated in Section 4, which is followed by a short discussion in Section 5. All technical proofs are given in the appendix.

\section{Linear methods for classification}
\subsection{Notations}
In this section, we review several popular classification tools from the aspect of dimension reduction. We consider a standard setup. Let $\cX=(\bX_1,\cdots,\bX_n)^{\top}$ be a $n\times p$ matrix with $n$ observations $\bX_1$,..., $\bX_n$, each of which is a $p$ dimensional vector. Let $\bY=(Y_1,...,Y_n)^{\top}$ be a response vector with $Y_i\in\{1,...,K\}$, $1\leq i\leq n$. That is, $\bX_i$ belongs to group $k$ if $Y_i=k$. Define the index set of group $k$ as $C_k=\{i: Y_i=k\}$ and its cardinality $n_k=|C_k|$, where $1\leq k\leq K$. The goal of classification is to establish a classification rule and label a new observation $\bX^*$ based on the training data.

In the literature, the Gaussian assumption is often used to facilitate statistical analysis of various methods. In the simplest setting, it is assumed that data from all groups share a common covariance matrix $\bSigma_w$, that is, $(\bX|Y=k)\sim \cN(\bmu_k,\bSigma_w)$, $1\leq k\leq K$. For easy presentation, we also assume that the prior probabilities $\pi_k=\bP(Y=k)$ are equal for all $k$. In practice, the prior probability can be always estimated and taken into account for most methods considered in this paper.

Recall that, for a specified strictly positive definite symmetric matrix $\bS$, the Mahalanobis distance between two vectors $\bu$ and $\bbv$ is
$$d_M(\bu,\bbv)=\left\{(\bu-\bbv)^{\top}\bS^{-1}(\bu-\bbv)\right\}^{\frac12}.$$
In fact, under the Gaussian assumption, the optimal classification rule minimizing the expected classification error is called the Bayes rule, which simply classifies a data point to a group with the nearest centroid in terms of Mahalanobis distance with $\bS=\bSigma_w$. That is,
\begin{eqnarray}\label{2.1}
Y=\argmin_{1\leq k\leq K}(\bX-\bmu_k)^{\top}\bSigma_w^{-1}(\bX-\bmu_k)=\argmin_{1\leq k\leq K}\left\|\bSigma_w^{-\frac12}(\bX-\bmu_k)\right\|^2_2.
\end{eqnarray}

A key observation from (\ref{2.1}) is that if we rotate the sample space by $\bSigma_w^{-1/2}$ first, then the Bayes rule is equivalent to a nearest centroid classifier with standard Euclidian distance.
Moreover, it follows (\ref{2.1}) that the optimal decision boundary separating groups $k$ and $\ell$ is an affine space defined by $\{\bX- (\bmu_k+\bmu_{\ell})/2\}\bSigma^{-1}_w(\bmu_k-\bmu_{\ell})=0$. Note that the normal vector to this affine space is  $\bSigma^{-1}_w(\bmu_k-\bmu_{\ell})$. Therefore, the decision boundary of the Bayes rule to the whole classification problem is a subset of the union of these affine spaces, whose normal vectors span a vector space $\bSigma^{-1}_w\bC$, where $\bC$ is the vector space spanned by $\{\bmu_k-\bmu_{\ell}\}_{1\leq k<\ell\leq K}$. Note that $\dim\bC=\dim (\spn\{\bmu_k-\bmu_K\}_{k=1}^{K-1})\leq K-1$, and the equal sign holds when the set of centroids $\{\bmu_k\}_{k=1}^K$ is in linear general position. By Lemma 2 in the appendix, when $p$ is larger than $K$, we will lose no information to project the data from $\mathbb{R}^p$ to a small subspace $\bSigma^{-1}_w\bC$ for classification. That is, it is equivalent to apply the Bayes rule to the projected data and the original data. In practice, when $p$ is large, it is extremely helpful to find a reasonable approximation subspace to $\bSigma^{-1}_w\bC$ to reduce the dimensionality.

Let us introduce some notations of important statistics. Without loss of generality, we assume that the features, i.e., columns of $\cX$, are centered to have mean zero since all methods considered in this paper are translation invariant. The within-class sample covariance matrix is
\[ \bW=\frac{1}{n}\sum_{k=1}^K\sum_{i\in C_k}(\bX_i-\hat\bmu_k)(\bX_i-\hat\bmu_k)^{\top},\]
where $\hat\bmu_k=n_k^{-1}\sum_{i\in C_k}\bX_i.$
The between-class sample covariance matrix is
\[\bB=\frac1n\sum_{k=1}^K n_k(\hat\bmu_k-\hat\bmu)(\hat\bmu_k-\hat\bmu)^{\top}=\frac1n\sum_{k=1}^K n_k\hat\bmu_k\hat\bmu_k^{\top},\] where $\hat\bmu=n^{-1}\sum_{k=1}^K n_k\hat\bmu_k=0$.
The total sample covariance matrix
\[\bT=n^{-1}\cX^{\top}\cX= \bW + \bB.\]

\subsection{Simple reduced-rank linear discriminant analysis}\label{s2.2}
A simple reduced-rank LDA \citep{hastie2009elements} projects the data to the centroid-spanning subspace
$\hat\bC=\spn \{\hat\bmu_k-\hat\bmu_{\ell}\}_{1\leq k<\ell\leq K}$. The idea is that, when calculating the (Euclidean) distances to find the closest centroid, one can ignore the distances orthogonal to $\hat\bC$ which contribute equally to all groups. This simple method reduces the dimensionality remarkably. The main drawback is that  it does not incorporate the covariance structure. We may lose much information if $\bSigma_w$ is far from a scalar matrix.

\subsection{Fisher's approach and the standard LDA}\label{s2.3}
Fisher's approach is to find a subspace so the projected centroids are spread out as much as possible with respect to the covariance. It finds the first direction by solving
\begin{eqnarray}
\bbv_1=\argmax_{\bbv}\frac{\bbv^{\top}\bB\bbv}{\bbv^{\top}\bW\bbv},
\end{eqnarray}
which is equivalent to
\begin{eqnarray}
\bbv_1=\argmax_{\bbv} {\bbv^{\top}\bB\bbv} \text{ subject to } {\bbv^{\top}\bW\bbv=1},
\end{eqnarray}
provided $\bW$ is not singular. It is well known that when $K=2$, $\bbv_1$ is the same as the normal vector (up to a scalar) of the decision boundary separating two groups obtained by standard LDA. When $K>2$, one can continue to solve this generalized eigenvalue problem until step $\rank(\bB)$, as follows.
\begin{eqnarray}
\bbv_2&=&\argmax_{\bbv} {\bbv^{\top}\bB\bbv} \text{ subject to } {\bbv^{\top}\bW\bbv=1; \text{ }\bbv^{\top}\bW\bbv_1=0},\nonumber \\
&\cdots&\nonumber\\
\bbv_k&=&\argmax_{\bbv} {\bbv^{\top}\bB\bbv} \text{ subject to } {\bbv^{\top}\bW\bbv=1; \text{ }\bbv^{\top}\bW\bbv_{\ell}=0},\text{ } \ell=1,2,...,k-1,\label{2.4}\\
&\cdots&\nonumber
\end{eqnarray}

Obviously, the covariance plays a role here through the sample pooled covariance $\bW$. Moreover, there is an order for these directions so the dimension of the subspace can be pre-specified or chosen data-adaptively. In fact, the simple reduced-rank LDA is a special case by restricting $\bW$ to a scalar matrix. In this sense, Fisher's approach is also a reduced-rank LDA method.

The standard LDA can be viewed as a dimension reduction technique as well. Roughly speaking, it mimics the Bayes rule by plugging in estimators of the common covariance and centroids. It labels an observation $\bX$ by $\hat Y=\argmin_{1\leq k\leq K} \|\bW^{-1/2}(\bX-\hat\bmu_k)\|_2$. Similar to the analysis of the Bayes rule, the normal vectors of the decision boundaries of standard LDA span a subspace $\bW^{-1}\hat\bC\subset\mathbb{R}^p$. It is equivalent to apply standard LDA to, instead of the original data, the projected data onto subspace $\bW^{-1}\hat\bC$. The following proposition indicates the equivalence between Fisher's approach and the standard LDA in the context of dimension reduction. It seems that this well-known result is not established formally anywhere so we include it here.

\begin{proposition}\label{p1}
If $\bW$ is nonsingular, then $\dim \bW^{-1}\hat\bC=\rank(\bB)$, and $\bW^{-1}\hat\bC=\spn\{\bbv_k\}_{k=1}^{r}$ where $r=\dim \hat\bC$, $\bbv_k$ is defined in (\ref{2.4}).
\end{proposition}

The original idea of the reduced-rank LDA is to find and project the data to a small subspace with high discriminant power before applying LDA. We see that, as a by product of the standard LDA, it gives automatically such a subspace, i.e., $\bW^{-1}\hat\bC$. A quick remark here is that Fisher's approach or the standard LDA performs well only when the sample size is large enough so $\bW^{-1}\hat\bC$ is a good approximation to $\bSigma^{-1}_w\bC$.

\subsection{The independence rule and related approaches}\label{s2.4}
It is well known that LDA does not work well when $p\sim n$ and $p>n$. In the context of dimension reduction reviewed in Section \ref{s2.3}, the reason is that $\bW^{-1}\hat\bC$ or $\bW^{-}\hat\bC$ is no longer a good approximation to $\bSigma_w^{-1}\bC$ for high dimension data, where $\bW^{-}$ is a pseudo-inverse of $\bW$. The main problem is that the sample pooled covariance $\bW$ is singular and a poor estimate for $\bSigma_w$. One remedy is to assume that the features are independent, i.e., $\bSigma_w$ is diagonal. This leads to the independence rule or diagonal LDA. To apply diagonal LDA, one just uses the diagonal part $\hat\bD_w=diag(\bW)$ instead of $\bW$ in the standard LDA. Although the features are rarely independent in applications, the IR or diagonal LDA usually outplays standard LDA when $p>n$ \citep{BickelLevina:2004}.

By the same spirit of Proposition \ref{p1}, one can see the equivalence between the IR and Fisher's approach with $\bW$ replaced by $\hat\bD_w$ in $(\ref{2.4})$, as stated in Corollary \ref{c2} in the appendix. \cite{witten2011penalized} further imposed some sparse assumptions on $\bbv_k$'s to derive a penalized LDA (PLDA). One can also conduct dimension reduction based on the rank of marginal discriminant power. Two well-known approaches are the NSC and FAIR, studied by \cite{tibshirani2002diagnosis} and \cite{FanFan:2008}, respectively.

\subsection{Principal component analysis}
As a dominant approach to dimension reduction in the context of unsupervised learning, PCA has been also used to solve supervised learning problems, e.g., principal component regression \citep{jolliffe2002principal}, supervised PCA \citep{bair2006prediction}, etc. Roughly speaking, PCA extracts the mutually orthogonal directions with largest variances of the data and discards the rest. In our context, standard PCA would ignore the label information and keep the eigenvectors corresponding to $q$ top eigenvalues of $\bT$, where $q$ can be pre-specified or chosen data adaptively. People have been concerned with this approach as there is no guarantee that the top principal components have good discriminant power. \cite{bair2006prediction} proposed a variant of supervised PCA. It is a two-stage procedure in which marginal statistics are used to reduce dimension before applying standard PCA. It seems that the label information is used only in the first stage. It is interesting to find alternative ways to conduct supervised PCA.

\subsection{Summary}
Reduced-rank LDA can be viewed as a general strategy which finds and projects the data into a linear subspace before applying standard LDA. Using Fisher's framework (\ref{2.4}) with different within-class covariance estimates, we unified simple reduced-rank LDA, standard LDA and IR as special cases. One main drawback of these methods is that none of them can compromise degeneracy of the sample covariance $\bW$ and non-triviality of the true covariance $\bSigma_w$. PCA would be a potentially good approach which requires neither nondegeneracy of $\bW$ nor independence assumption. Next, we will show a new reduced-rank LDA method which uses PCA in a supervised way.

\section{A New Reduced-Rank Linear Discriminant Analysis Method}

\subsection{Method}

To take advantage of existing methods described in the last section and study the multi-class classification problem in a unified manner, we consider $\bT_{\gamma}=\bW+\gamma\bB$ with $\gamma>0$. Recall that $\bW$ and $\bB$ are within class and between class sample covariance matrices, respectively. Here $\gamma$ is a tuning parameter. As we will see, if $\gamma=1$, our proposed procedure is equivalent to the standard PCA; and if $\gamma\to\infty$, the procedure is equivalent to the simple reduced-rank LDA illustrated in Section \ref{s2.2}.

Consider eigenvalue decomposition
\begin{eqnarray}\label{f3.1}
\bU_{\gamma}^{\top}\bT_{\gamma}\bU_{\gamma}= \bD_{\gamma}
\end{eqnarray}
where $\bD_{\gamma}$ is a diagonal matrix with diagonal entries ranked in a descending order and $\bU_{\gamma}$ is an orthogonal matrix. We propose a new reduced-rank LDA procedure based on the first $q$ principal components of $\bT_{\gamma}$. It can be conducted as follows.
\begin{enumerate}
\item Calculate $\bT_{\gamma}$ and $\bU_{\gamma}$ and project the data from $\mathbb{R}^p$ to the linear subspace spanned by the first $q$ columns of $\bU_{\gamma}$. In practice, the parameters $\gamma$ and $q$ can be chosen data adaptively.
\item Apply the standard LDA to the projected data.
\end{enumerate}

This procedure generalized classical methods such as PCA and simple reduced-rank LDA by allowing a varying parameter $\gamma$. The label information is taken into account through $\gamma$ in dimension reduction. In this sense, our procedure is a version of supervised PCA, so we call it supervised PCA-based LDA (SPCALDA). Although the parameter $\gamma$ plays a minor role in the later theoretical analysis, it makes our procedure more flexible. For example, the qualities of $\bW$ and $\bB$ to approximate their population counterparts are usually not equally good. Therefore, $\gamma$ can serve as a weight to balance them.

\subsection{Theoretical insights}\label{TI}
It is not uncommon that PCA is used to solve supervised learning problems, especially, in the analysis of high dimensional or highly correlated data. However, it is usually used as an ad-hoc method without any theoretical justification. We now illustrate some theoretical insights why our procedure (including PCA as a special case) works well under certain conditions.

To understand the proposed method, we consider a population version. Denote by $\bSigma_b$ and $\bSigma_t$ the population versions of between-class and total covariance matrix, respectively. Define $\bSigma_{\gamma}=\bSigma_w+\gamma \bSigma_b$, $\gamma>0$ with eigenvalue decomposition 
\[\bU_O^{\top}\bSigma_{\gamma}\bU_O=\bD_O,\]
where $\bD_O$ is a diagonal matrix with diagonal entries ranked in a descending order and $\bU_O$ is orthogonal. Because $\gamma$ plays only a minor role in this section, we drop it from notations $\bU_O$, $\bD_O$, etc. For an oracle procedure assuming parameters are known, we can project the data to the linear subspace spanned by top principal components of $\bU_O$. A natural question is when we can project the data by oracle procedure without information loss.

Let $\{\lambda_j\}_{j=1}^p$ be eigenvalues of $\bSigma_w$ in a descending order.  We consider a spiked covariance structure \cite{johnstone2001distribution}.

Spiked Condition: Assume that $\lambda_1\geq\cdots\geq\lambda_s>\lambda_{s+1}=\cdots=\lambda_p$ for some integer $s<p$.

\begin{theorem}\label{thm1}
If $p>s+K-1$, $s>1$, write $\bU_O=(\bU_{O1},\bU_{O2})$ where $\bU_{O1}$ and $\bU_{O2}$ are $p\times (s+K-1)$ and $p\times (p-s-K+1)$ matrices respectively. Under spiked condition, we have $$\bU^{\top}_{O2}\bSigma^{-1}_w(\bmu_k-\bmu_{\ell})=0, \text{ for all } 1\leq k<\ell\leq K.$$  In other words, $\bSigma^{-1}_w\bC$ are located in the subspace spanned by columns of $\bU_{O1}$.
\end{theorem}

In short, we will lose no discriminant power to project the data to a $s+K-1$ dimensional subspace spanned by columns of $\bU_{O1}$.

We may generalize this result further as follows. Without loss of generality, assume $\bmu=\sum_{k=1}^K\pi_k\bmu_k=0$. From the proof of Theorem \ref{thm1} in the appendix, we see the conclusion of Theorem \ref{thm1} holds for principal components of $\bSigma_{\brho}= \bSigma_w + \sum_{k=1}^{K}\rho_k \bmu_k \bmu_k^{\top},$ where $\brho=(\rho_1,...,\rho_{K})^{\top}$ with $\rho_k>0$, $1\leq k\leq K$.  Nevertheless, when $K$ is more than three, it is technically more complicated to tune $K$ parameters.

A more general model than the Gaussian one considered here is the mixture Gaussian model which allows each group to be distributed as mixture Gaussian with the same covariance; see e.g. \cite{hastie2009elements}, Section 12.7. It is useful in applications when the groups are inhomogeneous. Let $(\bX|Y=k)\sim \sum_{t=1}^{R_k}\pi_{kt}\cN(\bmu_{kt},\bSigma_w)$, where $1\leq k\leq K$, $1\leq t\leq R_k$, $\sum_{t=1}^{R_k}\pi_{kt}=1$. That is, there are $R_k$ prototypes for the group $k$. We can generalize Theorem \ref{thm1} to this case.

\begin{theorem}\label{thm2}
Let $R=\sum_{k=1}^K R_k$. Then Theorem 1 still holds if we replace $K$ by $R$ everywhere.
\end{theorem}

Remark 1. The spiked condition is crucial in Theorems 1 and 2. It is a reasonable model to approximate the true covariance when strong correlation among the features is present. It is also employed by \cite{HaoDongFan2015}, which aimed to sparsify the normal vector of optimal discriminant boundary for binary classification problems. In spite of the theoretical insight, the spiked condition plays no role in conducting our procedure. In real applications, it may not hold exactly. Nevertheless, our numerical studies show that our procedure performs very well.

Remark 2. In practice, we will work with $\bU$ instead of its population version $\bU_O$. Although $\bU_O$ might be very different from $\bU$ when $n\ll p$, the left part of $\bU$, i.e., $\bU_1$, can be similar to $\bU_{O1}$ under some conditions. For example, when the leading eigenvalues are large enough or their corresponding eigenvectors are sparse, $\bU_{O1}$ can be well-estimated by $\bU_1$ or its sparse counterpart \citep{johnstone2009consistency}.

\subsection{Computation}
In many contemporary applications, $p$ is much larger than $n$. For example, in some gene expression data sets, $p$ is a few thousands or more, and $n$ is a few hundreds or less. So it is time-consuming to calculate $p\times p$ matrix $\bT_{\gamma}$ and its eigenvalue decomposition directly. The following lemma offers a shortcut to find $\bU_1$.

\begin{lemma}\label{l1}
We can write $\bT_{\gamma}= n^{-1}\bA_{\gamma}^{\top}\bA_{\gamma}$, where $$\bA_{\gamma}=\left(\bX_1-\hat\bmu_{Y_1},...,\bX_n-\hat\bmu_{Y_n},(\gamma n_1)^{1/2}(\hat\bmu_1-\hat\bmu),...,(\gamma n_G)^{1/2}(\hat\bmu_K-\hat\bmu)\right)^{\top}$$
is an $(n+K)\times p$ matrix. Note that $\hat\bmu_{Y_i}=\hat\bmu_k$ when $Y_i=k$, $\hat\bmu=n^{-1}\sum_{k=1}^K n_k\hat\bmu_k=0$ by our convention.
\end{lemma}

When $p> n+K$ we can conduct eigenvalue decomposition for $(n+K)\times(n+K)$ matrix $\bA_{\gamma}\bA_{\gamma}^{\top}$ instead of $p\times p$ matrix $\bA_{\gamma}^{\top}\bA_{\gamma}$, because the eigenvectors of $\bA_{\gamma}^{\top}\bA_{\gamma}$ and $\bA_{\gamma}\bA_{\gamma}^{\top}$ are up to a linear transformation. To elaborate, by singular decomposition  $\bA_{\gamma}=\bV_{\gamma}\bGamma\bU_{\gamma}^{\top}$, where $\bGamma$ is diagonal, $\bV_{\gamma}$ is $(n+K)\times(n+K)$ orthogonal matrix, and $\bU_{\gamma}$ is $p\times p$ orthogonal matrix identical to $\bU_{\gamma}$ in (\ref{f3.1}). When $n+K$ is small or moderate, it is easy to find $\bV_{\gamma}$ which consists of eigenvectors of $\bA_{\gamma}\bA_{\gamma}^{\top}$. And the first $(n+K)$ columns of $\bU_{\gamma}$ can be obtained by standardizing $\bA_{\gamma}^{\top}\bV_{\gamma}$ column-wisely, as $\bA_{\gamma}^{\top}\bV_{\gamma}=\bU_{\gamma}\bGamma^{\top}$. In practice, it is sufficient to consider only the first $(n+K)$ columns of $\bU_{\gamma}$ (i.e., top principal components of $\bT_{\gamma}$), because the rest columns correspond to eigenvalue $0$ and contains little information.

For a fixed $K$, consider the scenario $n<p$. The computational complexities to find $\bA_{\gamma}\bA_{\gamma}^{\top}$ and conduct its singular decomposition are $O(n^2p)$ and $O(n^3)$, respectively. The computational complexity to find and standardize $\bA_{\gamma}^{\top}\bV_{\gamma}$ is $O(n^2p)$. Therefore, the overall computational complexity is $O(n^2p)$, which is linear with respect to $p$ for a fixed $n$. Therefore, our method is computationally efficient to analyze high dimensional data.

\section{Numerical studies}

\subsection{Simulated data examples}\label{s4.1}
We compare our proposed new reduced-rank LDA method (SPCALDA) with some other popular classification tools by simulated data examples. In particular, we considered simple reduced-rank LDA (SRRLDA) reviewed in Section \ref{s2.2}, LDA after standard PCA (PCALDA) that is a special case of SPCALDA with fixed $\gamma=1$, and the independence rule (IR) reviewed in Section \ref{s2.4}. Moreover, we added the Bayes rule as an oracle benchmark for comparison.

Six scenarios are considered here. For each scenario, 200 observations are generated and equally split between the four classes. Among 200 observations, 100 are assigned to the training set, and the rest 100 serve as test data. There are $p=500$ features. For each class $k$, $\bX\sim\cN(\bmu_k,\bSigma_w)$, where $\bmu_k$ and $\bSigma_w$ are defined as follows.
\begin{description}
  \item[Scenario 1.] The common covariance $\bSigma_w=\bI_p$. The mean vectors are given by $\mu_{1j}=0.3*\mathbb{I}_{1\leq j\leq 125}$, $\mu_{2j}=0.3*\mathbb{I}_{126\leq j\leq 250}$, $\mu_{3j}=0.3*\mathbb{I}_{251\leq j\leq 375}$, $\mu_{4j}=0.3*\mathbb{I}_{376\leq j\leq 500}$, where $\mathbb{I}_{\cS}$ is a vector with entries 1 for indices in $\cS$ and 0 elsewhere.
  \item[Scenario 2.] $\bSigma_w=\bI_p$. $\mu_{1j}\sim \cN(0,0.3^2)$ when $1\leq j\leq 125$, and $\mu_{1j}=0$ otherwise, $\mu_{2j}\sim \cN(0,0.3^2)$ when $126\leq j\leq 250$, and $\mu_{2j}=0$ otherwise, $\mu_{3j}\sim \cN(0,0.3^2)$ when $251\leq j\leq 375$, and $\mu_{3j}=0$ otherwise, $\mu_{4j}\sim \cN(0,0.3^2)$ when $376\leq j\leq 500$, and $\mu_{4j}=0$ otherwise.
  \item[Scenario 3.] $\bSigma_w=(\sigma_{ij})$ with $\sigma_{ii}=1$ and $\sigma_{ij}=0.5$ for $i\ne j$. $\mu_{1j}=0.21*\mathbb{I}_{1\leq j\leq 125}$, $\mu_{2j}=0.21*\mathbb{I}_{126\leq j\leq 250}$, $\mu_{3j}=0.21*\mathbb{I}_{251\leq j\leq 375}$, $\mu_{4j}=0.21*\mathbb{I}_{376\leq j\leq 500}$.
  \item[Scenario 4.] $\bSigma_w$ is the same as in Scenario 3. $\mu_{1j}\sim \cN(0,0.21^2)$ when $1\leq j\leq 125$, and $\mu_{1j}=0$ otherwise, $\mu_{2j}\sim \cN(0,0.21^2)$ when $126\leq j\leq 250$, and $\mu_{2j}=0$ otherwise, $\mu_{3j}\sim \cN(0,0.21^2)$ when $251\leq j\leq 375$, and $\mu_{3j}=0$ otherwise, $\mu_{4j}\sim \cN(0,0.21^2)$ when $376\leq j\leq 500$, and $\mu_{4j}=0$ otherwise.
\end{description}

To test the robustness of the proposed method against departures from Gaussian and equal covariance assumptions, we include two more scenarios. In particular, Scenario 5 considers a case when the data are contaminated by a random heavy-tailed noise. Scenario 6 considers the situation when the observations from different classes do not share a common covariance structure.
\begin{description}
 \item[Scenario 5.] $\bX$ is generated as in Scenario 3. Let $\bZ$ is a $p$ dimensional random vector with entries IID from $t$-distribution with degrees of freedom 3. We assume that realizations of $\tilde \bX=\bX+0.2\bZ$ are observed instead of $\bX$.
 \item[Scenario 6.] $\bX$ is generated as in Scenario 3. For each class $k$, we first generate a $p$ dimensional vector $\bd_k$ with entries IID from standard uniform distribution, and define a diagonal covariance matrix $\bDelta_k=$diag$(\bd_k^2)$. For each class $k$, let $\bZ\sim\cN(\bzero,\bDelta_k)$. We assume that realizations of $\tilde \bX=\bX+\bZ$ are observed instead of $\bX$.
\end{description}

There are two tuning parameters for SPCALDA (i.e., $\gamma$ and $q$) and one parameter for PCALDA method, which were chosen by five-fold cross validation. 
The fitted models were evaluated using the test set for all methods. We repeated each experiment 100 times. The average and standard deviation of classification error rates for each method are listed in Table \ref{Table1}. The SPCALDA method always outperformed PCALDA, which indicates that it is helpful to tune the parameter $\gamma$. In the independence cases, SPCALDA is comparable with SRRLDA and IR but much better than them in the correlated cases. In general, the performance of SRRLDA and IR highly depends on the true covariance structure and SPCALDA is robust over different covariance structures. Moreover, when the Gaussian assumption or equal covariance assumption is violated, we see that SPCALDA still performs reasonably well.

 \begin{table}[htp]
\caption{Mean (and standard errors) of classification error rates (\%).}\label{Table1}
\centering
\begin{tabular}{|c|ccccc|}
\hline
         &SPCALDA & PCALDA   &SRRLDA   &IR    &Oracle\\
\hline
Scenario 1&18.93(4)&26.53(4.52)&19.33(3.94)&18.45(3.86)&2.69(1.6)\\
Scenario 2&19.96(3.91)&27.71(5.1)&20.46(4.7)&19.29(4.03)&2.8(1.75)\\
Scenario 3&20.73(4.32)&30(5.64)&36.61(10.75)&63.92(5.41)&2.73(1.63)\\
Scenario 4&22.78(4.4)&32.26(5.82)&38.61(10.31)&64.38(7.92)&3.07(1.69)\\
Scenario 5&28.8(4.82)&38.42(6.41)&43.52(9.66)&64.38(5.8)&NA\\
Scenario 6&38.29(5.35)&50.75(6.72)&49.44(8.85)&64.79(6.57)&NA\\
 \hline
\end{tabular}
\end{table}

\subsection{Real data examples}
In this section, we illustrate the performance of our method using six popular gene expression data sets, which have been studied in the literature. In particular, we considered three binary data sets, \textbf{Chin} \citep{chin2006genomic}, \textbf{Chowdary} \citep{chowdary2006prognostic}, \textbf{Gordon} \citep{gordon2002translation}, and three multi-class data sets, \textbf{Golub} \citep{golub1999molecular}, \textbf{Nakayama} \citep{nakayama2007gene}, and \textbf{Sun} \citep{sun2006neuronal}. The three binary data sets are available in R package \texttt{datamicroarray}. The data set \textbf{Golub} is available in R package \texttt{golubEsets}. The original \textbf{Nakayama} data set contains 105 samples from 10 types of soft tissue tumors. We considered a subset of 86 samples belonging to 5 tumor types and ignored the rest tumor types for which less than 7 samples were available. \textbf{Nakayama} and \textbf{Sun} are available on Gene Expression Omnibus \citep{barrett2005ncbi} with accession numbers GDS2736 and GDS1962, respectively. We list in Table \ref{Table1.5} the sample size, number of features, number of classes, data distribution among different classes and related disease for each data set.

\begin{table}[htp]
\caption{Data sets used in this study.}\label{Table1.5}
\centering
\begin{tabular}{|c|ccccc|}
\hline
Data set          &related disease & \# samples & \# features & \# classes & data distribution\\
\hline
\textbf{Chin}     & breast cancer   & 118      & 22,215      & 2          &43, 75\\
\textbf{Chowdary} & breast cancer   & 104      & 22,283      & 2          & 42, 62\\
\textbf{Gordon}   & lung cancer     & 181      & 12,533      & 2          & 87, 94\\
\textbf{Golub}    & leukemia        & 72       & 7,129       & 3          & 9, 25, 38\\
\textbf{Nakayama} & soft tissue tumor& 86      & 22,283      & 5          &15, 15, 16, 19, 21\\
\textbf{Sun}      & glioma          & 180      & 54,613      & 4          &23, 26, 50, 81\\
 \hline
\end{tabular}
\end{table}

Besides the methods considered in Section \ref{s4.1}, we included some state-of-the-art multi-class classification tools including NSC \citep{tibshirani2002diagnosis}, RDA \citep{guo2007regularized}, PLDA \citep{witten2011penalized}, and SDA \citep{clemmensen2012sparse}, which have been implemented by \texttt{R} packages \texttt{pamr}, \texttt{rda}, \texttt{penalizedLDA}, and \texttt{sparseLDA}, respectively. These methods are based on various sparsity assumptions.

For each of these data sets, we randomly split the data with a 3 to 1 ratio into a training set and a test set, in a balanced manner. Five-fold cross-validation was conducted on the training set to select the tuning parameters for all methods, and the classification error rates using test set were recorded. In Table \ref{Table2} we list the average classification error rates and their standard deviations over 25 random training/test set splits. We omit the results of PCALDA and IR which are dominated by SPCALDA and NSC, respectively. We see that SPCALDA performed best for two data sets, and second best for four data sets. In particular, SPCALDA is clearly the best in pairwise comparisons with other methods. We list in Table \ref{Table3} the computation time for each method. We find that all these methods are reasonably fast to handle contemporary high dimensional data sets. In summary, our proposed method SPCALDA offers a competitive classification tool for high dimensional gene expression data.

\begin{table}[htp]
\caption{Mean (and standard errors) of classification error rates (\%).}\label{Table2}
\centering
\begin{tabular}{|c|cccccc|}
\hline
 &SPCALDA&SRRLDA&NSC&PLDA&RDA&SDA\\
\hline
\textbf{Chin} &11.57(6.57)&12.11(6.13)&12.41(7.42)&13.87(6.79)&12.25(5.32)&10.13(4.47)\\
\textbf{Chowdary}&4.13(3.62)&10.43(5.82)&5.19(5.03)&33.63(9.52)&4.75(3.9)&17.19(8.26)\\
\textbf{Gordon}&0.62(1.19)&2.29(2.99)&0.79(1.08)&0.53(0.96)&1.4(1.25)&6.02(3.41)\\
\textbf{Golub}&5.43(5.44)&24.2(12.93)&4.6(4.17)&7.41(5.91)&6.3(3.74)&15.89(11.62)\\
\textbf{Nakayama}&16.37(7.05)&20.6(8.94)&23.51(6.36)&27.6(8.84)&15.68(7.98)&33.73(6.89)\\
\textbf{Sun}&30.43(5.73)&31.63(6.89)&33.24(6.03)&33.21(5.89)&33.48(6.97)&33.33(8.78)\\
\hline
\end{tabular}
\end{table}

\begin{table}[htp]
\caption{Mean (and standard error) of computation time per replicate (in second).}\label{Table3}
\centering
\begin{tabular}{|c|cccccc|}
\hline
 &SPCA-LDA&SRR-LDA&NSC&PLDA&RDA&SDA\\
\hline
\textbf{Chin} &14.5(1.03)&0.05(0.03)&3.93(0.61)&13.51(0.77)&51.17(3.27)&1.12(0.22)\\
\textbf{Chowdary}&12.09(0.25)&0.05(0.04)&3.63(0.21)&12.41(0.75)&49.52(0.43)&1.09(0.17)\\
\textbf{Gordon}&11.68(0.15)&0.04(0.02)&1.95(0.07)&6.82(0.1)&30.2(0.44)&0.53(0.04)\\
\textbf{Golub}&2.98(0.08)&0.01(0.01)&0.71(0.04)&3.9(0.34)&10.97(0.25)&0.22(0.03)\\
\textbf{Nakayama}&8.77(0.2)&0.06(0.02)&2.33(0.09)&27.46(0.73)&38.14(0.55)&1.75(0.22)\\
\textbf{Sun}&55.16(1.19)&0.24(0.02)&9.16(0.28)&56.3(0.94)&162.03(2.46)&6.42(0.75)\\
\hline
\end{tabular}
\end{table}

\section{Discussion}
Feature selection and feature extraction are two popular strategies in statistical machine learning. In the context of this paper, the sparse methods such as NSC and SDA can conduct variable selection and model estimation simultaneously, and belong to the first category. On the other hand, our methods, including classical reduced-rank LDA and PCA as special cases, belong to the latter. Both of these two approaches have their strength and weakness. For example, sparse methods can achieve model selection consistency and efficiency under various sparse assumptions. But they may fail when the true model is far from sparse. In contrast, our approach does not depend on sparse assumptions and is robust against the sparsity level of the true model. Our real data examples also verify the robustness of our method. In general, we can not expect a result on model selection consistency or efficiency. Nevertheless, we discuss a spiked covariance condition under which our method may achieve efficiency.


\section*{Acknowledgements}
This research was partially supported by grants DMS 1309507 from National Science Foundation and 11671022 from National Science Foundation of China. The authors are grateful to Daniela Witten for helpful discussion, and to the editor, associate editor, and two referees for comments and suggestions that greatly improved the paper.

\section{Appendix}
\textbf{Proof of Proposition \ref{p1}.} Recall that, by our convention, the data have been centered, $\hat\bmu=n^{-1}\sum_{k=1}^K n_k\hat\bmu_k=0$, so $\bB=n^{-1}\sum_{k=1}^K n_k\hat\bmu_k\hat\bmu_k^{\top}$. Note that $\bB$ is semi-positive definite.

For a special case $\bW=\bI$, $\{\bbv_k\}_{k=1}^r$ are just eigenvectors of $\bB$ corresponding to positive eigenvalues. For any vector $\bu\perp \hat\bC$, we have
\begin{eqnarray*}
& & \bu\perp\hat\bmu_k, \quad k=1,2,...,K\\
&\Leftrightarrow & \bu^{\top}\bB\bu=\frac1n\sum_{k=1}^K n_k\bu^{\top}\hat\bmu_k\hat\bmu_k^{\top}\bu=\frac1n\sum_{k=1}^K n_k (\hat\bmu_k^{\top}\bu)^2=0\\
&\Leftrightarrow & \bu \text{ belongs to the eigen-space of } \bB\text{ corresponding to eigenvalue }0 \\
&\Leftrightarrow & \bu\perp\spn\{\bbv_k\}_{k=1}^r.
\end{eqnarray*}

That is, $\hat\bC$ and $\spn\{\bbv_k\}_{k=1}^r$ have the same orthogonal complement. Hence they are the same linear subspace and have the same dimension.

For arbitrary nonsingular $\bW$, we may transform the data by linear operator $\bW^{-1/2}$. That is, define $\tilde\bX_i=\bW^{-1/2}\bX_i$, $1\leq i\leq n$. It is easy to see that the statistics after transformation satisfy $\tilde\bW=\bI$, $\tilde\bB=\bW^{-1/2}\bB\bW^{-1/2}$, $\tilde{\bmu}_k=\bW^{-1/2}\hat\bmu_k$, $\tilde{\bC}=\bW^{-1/2}\hat\bC$, $\tilde\bbv_k=\bW^{1/2}\bbv_k$ (no negative sign on the power). By the argument above, we have $\tilde{\bC}=\spn\{\tilde \bbv_k\}_{k=1}^r$, so $\bW^{-1}\hat\bC=\bW^{-1/2}\tilde{\bC}=\spn\{\bW^{-1/2}\tilde \bbv_k\}_{k=1}^r=\spn\{\bbv_k\}_{k=1}^r$.

In fact, the proof goes through if $\bW$ is replaced by an arbitrary nonsingular equivariant covariance estimator. Hence we have the following corollary.

\begin{corollary}\label{c2}
The conclusion of Proposition \ref{p1} still holds if $\bW$ is replaced by any nonsingular equivariant within-class covariance estimate. In particular, replacing $\bW$ by its diagonal part $\hat\bD_w$, we can view diagonal LDA as a dimension reduction tool.
\end{corollary}

\textbf{Proof of Theorem \ref{thm1}.} We show a proof for a large family described in Remark 5 $\bSigma_{\brho}= \bSigma_w + \sum_{k=1}^{K}\rho_k \bmu_k \bmu_k^{\top},$ where $\brho=(\rho_1,...,\rho_{K})^{\top}$ with $\rho_k>0$ for all $k$. Theorem 1 can be obtained as a special case because the family $\{\bSigma_{\gamma}\}_{\gamma>0}$ is included in the larger one.

Let us fix an arbitrary $\brho=(\rho_1,...,\rho_{K})^{\top}$ with all positive entries, and $\bU_O^{\top}\bSigma_{\brho}\bU_O=\bD_O$. By the spiked condition, we can write
\[\bSigma_w= \lambda_p\bI+\sum_{i=1}^s(\lambda_{i}-\lambda_p)\bxi_{i}\bxi_{i}^{\top},\]
where $\{\xi_{i}\}_{i=1}^s$ are eigenvectors to eigenvalues larger than $\lambda_p$.
For $1\leq k<\ell\leq K$, we have
\begin{eqnarray}
& & \bSigma^{-1}_w(\bmu_k-\bmu_{\ell})\nonumber \\
&=& \left(\lambda_p\bI+\sum_{i=1}^s(\lambda_{i}-\lambda_p)\bxi_{i}\bxi_{i}^{\top}\right)^{-1}(\bmu_k-\bmu_{\ell})\nonumber\\
&=& \left(\lambda_p^{-1}\bI-\sum_{i=1}^s\frac{\lambda_{i}-\lambda_p}{\lambda_p\lambda_i}\bxi_{i}\bxi_{i}^{\top}\right)(\bmu_k-\bmu_{\ell})\nonumber\\
&=& \lambda_p^{-1}(\bmu_k-\bmu_{\ell})-\sum_{i=1}^s\left[\frac{\lambda_{i}-\lambda_p}{\lambda_p\lambda_i}\bxi_{i}^{\top}(\bmu_k-\bmu_{\ell})\right]\bxi_{i}\nonumber\\
&\in& \spn\{\bmu_k-\bmu_{\ell},\bxi_1,...,\bxi_s\}\label{f6.1}.
\end{eqnarray}

Moreover,
\begin{eqnarray}\label{f6.2}
\bSigma_{\brho}= \lambda_p\bI+\sum_{i=1}^s(\lambda_{i}-\lambda_p)\bxi_{i}\bxi_{i}^{\top} + \sum_{k=1}^{K}\rho_k \bmu_k \bmu_k^{\top}.
\end{eqnarray}
If $p>s+K-1$, the dimension of linear subspace $\bS = \spn\left\{\{\bxi_i\}_{i=1}^s,\{\bmu_k\}_{k=1}^K\right\}$ is at most $s+K-1$ because of our convention $\sum_{k=1}^K\pi_k\bmu_k=0$. On one hand,
by (\ref{f6.1}), $\bSigma^{-1}_w(\bmu_k-\bmu_{\ell})\in \bS$. On other the hand, the eigenspace of $\bSigma_{\brho}$ corresponding to eigenvalue $\lambda_p$ is orthogonal to $\bS$ by (\ref{f6.2}). Therefore, columns of $\bU_{O2}$ are orthogonal to $\bS$, and hence to $\bSigma^{-1}_w(\bmu_k-\bmu_{\ell})$ for all $k$, $\ell$.

\textbf{Proof of Theorem \ref{thm2}.} The proof follows the proof of Theorem \ref{thm1} by noticing that $\bmu_k=\sum_{t=1}^{R_k}\pi_{kt}\bmu_{kt}$, and $\spn\{\bmu_k\}_{k=1}^K\subset\spn\{\bmu_{kt}:1\leq k\leq K;1\leq t\leq R_k\}$.

\textbf{Proof of Lemma \ref{l1}.}
\begin{eqnarray*}
\bT_{\gamma}&=&\bW+\gamma\bB\\
&=&\frac1n\left(\sum_{k=1}^K\sum_{i\in C_k}(\bX_i-\hat\bmu_k)(\bX_i-\hat\bmu_k)^{\top}+\sum_{k=1}^K \gamma n_k(\hat\bmu_k-\hat\bmu)(\hat\bmu_k-\hat\bmu)^{\top}\right) \\
&=&\frac1n\left(\sum_{i=1}^i(\bX_i-\hat\bmu_{Y_i})(\bX_i-\hat\bmu_{Y_i})^{\top}+\sum_{k=1}^K \gamma n_k(\hat\bmu_k-\hat\bmu)(\hat\bmu_k-\hat\bmu)^{\top}\right) \\
&=&\frac{1}{n}\bA_{\gamma}^{\top}\bA_{\gamma}
\end{eqnarray*}

\begin{lemma}\label{l2}
In the context of formula (\ref{2.1}), let $\bbeta_{k,\ell}=\bSigma^{-1}_w(\bmu_k-\bmu_{\ell})$ and $\bH\subset \mathbb{R}^p$ is arbitrary linear subspace such as $\bbeta_{k,\ell}\in\bH$.
Let $\bP_{\bH}$ be the projection operator from $\mathbb{R}^p$ to $\bH$. Then the normal vector to the optimal discriminant boundary separating groups $k$ and $\ell$ using information from only the projected data $\bP_{\bH}(\bX)$ is the same as $\bbeta_{k,\ell}$.
\end{lemma}


The conclusion below (\ref{2.1}) follows Lemma \ref{l2} with the choice $\bH=\bSigma_w^{-1}\bC$.

\textbf{Proof of Lemma \ref{l2}.}
Let $\{\bh_j\}_{j=1}^p$ be an orthonormal basis for $\mathbb{R}^p$, and $\bH=\spn\{\bh_j\}_{j=1}^q$, $\bG=\spn\{\bh_j\}_{j=q+1}^p$. By abuse of notation, we also use $\bH$ and $\bG$ to denote $q\times p$ matrix $(\bh_1,...,\bh_q)^{\top}$ and $(p-q)\times p$ matrix $(\bh_{q+1},...,\bh_p)^{\top}$, respectively. Let $\bF=(\bH^{\top},\bG^{\top})^{\top}$ be an orthogonal matrix. Let $\tilde \bX=\bF\bX$. Then $(\tilde \bX|Y=k)\sim \cN(\bF\bmu_k,\bF\bSigma_w\bF^{\top})$.

Now we work on an equivalent model $(\tilde \bX,Y)$, where the projection $\bP_{\bH}$ is simply a projection to the first $q$ coordinates. In this equivalent model, it is sufficient to show that the optimal discriminant boundaries obtained from whole data $\tilde \bX$ and the projected data are exactly the same.

First, using the whole data $\tilde\bX$, the normal vector to the optimal discriminant boundary separating groups $k$ and $\ell$ is
\begin{eqnarray}\label{6.3}
\tilde \bbeta_{k,\ell}= \left(\bF\bSigma_w\bF^{\top}\right)^{-1}(\bF\bmu_k-\bF\bmu_{\ell})=\bF \bSigma_w^{-1}(\bmu_k-\bmu_{\ell})=\bF\bbeta_{k,\ell}.
\end{eqnarray}
Note that the condition $\bbeta_{k,\ell}\in\bH$ implies $\bF\bbeta_{k,\ell}= {\bH\bbeta_{k,\ell} \choose \bG\bbeta_{k,\ell}} = {\bH\bbeta_{k,\ell} \choose \bzero} $. That is, $\tilde \bbeta_{k,\ell}$ is a sparse vector supported in its first $q$ coordinates. By (\ref{6.3}), we have
\[ \bF(\bmu_k-\bmu_{\ell})=\left(\bF\bSigma_w\bF^{\top}\right)\bF\bbeta_{k,\ell},\] which implies
\[ {\bH(\bmu_k-\bmu_{\ell})\choose\bG(\bmu_k-\bmu_{\ell})}=\left(
                                                                                   \begin{array}{cc}
                                                                                     \bH\bSigma_w\bH^{\top} & \bG\bSigma_w\bH^{\top} \\
                                                                                     \bH\bSigma_w\bG^{\top} & \bG\bSigma_w\bG^{\top} \\
                                                                                   \end{array}
                                                                                 \right){\bH\bbeta_{k,\ell} \choose \bzero}.
\]
Comparing the top $q$ rows of both sides, we have $\bH(\bmu_k-\bmu_{\ell})=(\bH\bSigma_w\bH^{\top})\bH\bbeta_{k,\ell}$. So
\begin{eqnarray}\label{6.4}
\bH\bbeta_{k,\ell} =(\bH\bSigma_w\bH^{\top})^{-1}\bH(\bmu_k-\bmu_{\ell}).
\end{eqnarray} 
To summarise, $\tilde \bbeta_{k,\ell}$ is a sparse vector with its first $q$ coordinates defined as in (\ref{6.4}).

Second, we consider the projected data. Write $\tilde \bX={\bH\bX\choose\bG\bX}={\tilde \bX_1\choose\tilde \bX_2}$, where $\tilde\bX_1|Y=k\sim\cN(\bH\bmu_k,\bH\bSigma_w\bH^{\top})$. Using information from the projected data $\tilde\bX_1$ only, we find the normal vector to the optimal discriminant boundary is $(\bH\bSigma_w\bH^{\top})^{-1}\bH(\bmu_k-\bmu_{\ell})$ which is the same as $\bH\bbeta_{k,\ell} $ by (\ref{6.4}). Therefore, we lose no information to retain $\tilde \bbeta_{k,\ell}$ using projected data $\tilde\bX_1$ instead of whole data $\tilde\bX$.

\bibliographystyle{biometrika}
\bibliography{SPCA}

\begin{thebibliography}{23}
\expandafter\ifx\csname natexlab\endcsname\relax\def\natexlab#1{#1}\fi

\bibitem[{Bair et~al.(2006)Bair, Hastie, Paul \&
  Tibshirani}]{bair2006prediction}
\textsc{Bair, E.}, \textsc{Hastie, T.}, \textsc{Paul, D.} \&
  \textsc{Tibshirani, R.} (2006).
\newblock Prediction by supervised principal components.
\newblock \textit{Journal of the American Statistical Association}
  \textbf{101}.

\bibitem[{Barrett et~al.(2005)Barrett, Suzek, Troup, Wilhite, Ngau, Ledoux,
  Rudnev, Lash, Fujibuchi \& Edgar}]{barrett2005ncbi}
\textsc{Barrett, T.}, \textsc{Suzek, T.~O.}, \textsc{Troup, D.~B.},
  \textsc{Wilhite, S.~E.}, \textsc{Ngau, W.-C.}, \textsc{Ledoux, P.},
  \textsc{Rudnev, D.}, \textsc{Lash, A.~E.}, \textsc{Fujibuchi, W.} \&
  \textsc{Edgar, R.} (2005).
\newblock Ncbi geo: mining millions of expression profiles—database and tools.
\newblock \textit{Nucleic acids research} \textbf{33}, D562--D566.

\bibitem[{Bickel \& Levina(2004)}]{BickelLevina:2004}
\textsc{Bickel, P.} \& \textsc{Levina, E.} (2004).
\newblock Some theory for fisher's linear discriminant function,naive bayes',
  and some alternatives when there are many more variables than observations.
\newblock \textit{Bernoulli} \textbf{10}, 989--1010.

\bibitem[{Cai \& Liu(2011)}]{cai2011direct}
\textsc{Cai, T.} \& \textsc{Liu, W.} (2011).
\newblock A direct estimation approach to sparse linear discriminant analysis.
\newblock \textit{Journal of the American Statistical Association}
  \textbf{106}.

\bibitem[{Chin et~al.(2006)Chin, DeVries, Fridlyand, Spellman, Roydasgupta,
  Kuo, Lapuk, Neve, Qian, Ryder et~al.}]{chin2006genomic}
\textsc{Chin, K.}, \textsc{DeVries, S.}, \textsc{Fridlyand, J.},
  \textsc{Spellman, P.~T.}, \textsc{Roydasgupta, R.}, \textsc{Kuo, W.-L.},
  \textsc{Lapuk, A.}, \textsc{Neve, R.~M.}, \textsc{Qian, Z.}, \textsc{Ryder,
  T.} et~al. (2006).
\newblock Genomic and transcriptional aberrations linked to breast cancer
  pathophysiologies.
\newblock \textit{Cancer cell} \textbf{10}, 529--541.

\bibitem[{Chowdary et~al.(2006)Chowdary, Lathrop, Skelton, Curtin, Briggs,
  Zhang, Yu, Wang \& Mazumder}]{chowdary2006prognostic}
\textsc{Chowdary, D.}, \textsc{Lathrop, J.}, \textsc{Skelton, J.},
  \textsc{Curtin, K.}, \textsc{Briggs, T.}, \textsc{Zhang, Y.}, \textsc{Yu,
  J.}, \textsc{Wang, Y.} \& \textsc{Mazumder, A.} (2006).
\newblock Prognostic gene expression signatures can be measured in tissues
  collected in rnalater preservative.
\newblock \textit{The journal of molecular diagnostics} \textbf{8}, 31--39.

\bibitem[{Clemmensen et~al.(2012)Clemmensen, Hastie, Witten \&
  Ersb{\o}ll}]{clemmensen2012sparse}
\textsc{Clemmensen, L.}, \textsc{Hastie, T.}, \textsc{Witten, D.} \&
  \textsc{Ersb{\o}ll, B.} (2012).
\newblock Sparse discriminant analysis.
\newblock \textit{Technometrics} .

\bibitem[{Fan \& Fan(2008)}]{FanFan:2008}
\textsc{Fan, J.} \& \textsc{Fan, Y.} (2008).
\newblock High dimensional classification using features annealed independence
  rules.
\newblock \textit{Annals of statistics} \textbf{36}, 2605.

\bibitem[{Fan et~al.(2012)Fan, Feng \& Tong}]{fan2012road}
\textsc{Fan, J.}, \textsc{Feng, Y.} \& \textsc{Tong, X.} (2012).
\newblock A road to classification in high dimensional space: the regularized
  optimal affine discriminant.
\newblock \textit{Journal of the Royal Statistical Society: Series B
  (Statistical Methodology)} \textbf{74}, 745--771.

\bibitem[{Golub et~al.(1999)Golub, Slonim, Tamayo, Huard, Gaasenbeek, Mesirov,
  Coller, Loh, Downing, Caligiuri et~al.}]{golub1999molecular}
\textsc{Golub, T.~R.}, \textsc{Slonim, D.~K.}, \textsc{Tamayo, P.},
  \textsc{Huard, C.}, \textsc{Gaasenbeek, M.}, \textsc{Mesirov, J.~P.},
  \textsc{Coller, H.}, \textsc{Loh, M.~L.}, \textsc{Downing, J.~R.},
  \textsc{Caligiuri, M.~A.} et~al. (1999).
\newblock Molecular classification of cancer: class discovery and class
  prediction by gene expression monitoring.
\newblock \textit{science} \textbf{286}, 531--537.

\bibitem[{Gordon et~al.(2002)Gordon, Jensen, Hsiao, Gullans, Blumenstock,
  Ramaswamy, Richards, Sugarbaker \& Bueno}]{gordon2002translation}
\textsc{Gordon, G.~J.}, \textsc{Jensen, R.~V.}, \textsc{Hsiao, L.-L.},
  \textsc{Gullans, S.~R.}, \textsc{Blumenstock, J.~E.}, \textsc{Ramaswamy, S.},
  \textsc{Richards, W.~G.}, \textsc{Sugarbaker, D.~J.} \& \textsc{Bueno, R.}
  (2002).
\newblock Translation of microarray data into clinically relevant cancer
  diagnostic tests using gene expression ratios in lung cancer and
  mesothelioma.
\newblock \textit{Cancer research} \textbf{62}, 4963--4967.

\bibitem[{Guo et~al.(2007)Guo, Hastie \& Tibshirani}]{guo2007regularized}
\textsc{Guo, Y.}, \textsc{Hastie, T.} \& \textsc{Tibshirani, R.} (2007).
\newblock Regularized linear discriminant analysis and its application in
  microarrays.
\newblock \textit{Biostatistics} \textbf{8}, 86--100.

\bibitem[{Hao et~al.(2015)Hao, Dong \& Fan}]{HaoDongFan2015}
\textsc{Hao, N.}, \textsc{Dong, B.} \& \textsc{Fan, J.} (2015).
\newblock Sparsifying the fisher linear discriminant by rotation.
\newblock \textit{Journal of the Royal Statistical Society: Series B
  (Statistical Methodology)} .

\bibitem[{Hastie et~al.(2009)Hastie, Tibshirani \&
  Friedman}]{hastie2009elements}
\textsc{Hastie, T.}, \textsc{Tibshirani, R.} \& \textsc{Friedman, J.} (2009).
\newblock \textit{The Elements of Statistical Learning: Data Mining, Inference,
  and Prediction}.
\newblock Springer series in statistics. Springer-Verlag New York.

\bibitem[{Johnstone(2001)}]{johnstone2001distribution}
\textsc{Johnstone, I.} (2001).
\newblock On the distribution of the largest eigenvalue in principal components
  analysis.
\newblock \textit{Ann. Statist} \textbf{29}, 295--327.

\bibitem[{Johnstone \& Lu(2009)}]{johnstone2009consistency}
\textsc{Johnstone, I.} \& \textsc{Lu, A.} (2009).
\newblock On consistency and sparsity for principal components analysis in high
  dimensions.
\newblock \textit{Journal of the American Statistical Association}
  \textbf{104}, 682--693.

\bibitem[{Jolliffe(2002)}]{jolliffe2002principal}
\textsc{Jolliffe, I.} (2002).
\newblock \textit{Principal Component Analysis}.
\newblock Springer Series in Statistics. Springer.

\bibitem[{Mai et~al.(2015)Mai, Yang \& Zou}]{mai2015multiclass}
\textsc{Mai, Q.}, \textsc{Yang, Y.} \& \textsc{Zou, H.} (2015).
\newblock Multiclass sparse discriminant analysis.
\newblock \textit{arXiv preprint arXiv:1504.05845} .

\bibitem[{Mai et~al.(2012)Mai, Zou \& Yuan}]{mai2012direct}
\textsc{Mai, Q.}, \textsc{Zou, H.} \& \textsc{Yuan, M.} (2012).
\newblock A direct approach to sparse discriminant analysis in ultra-high
  dimensions.
\newblock \textit{Biometrika} \textbf{102}, 33--45.

\bibitem[{Nakayama et~al.(2007)Nakayama, Nemoto, Takahashi, Ohta, Kawai, Seki,
  Yoshida, Toyama, Ichikawa \& Hasegawa}]{nakayama2007gene}
\textsc{Nakayama, R.}, \textsc{Nemoto, T.}, \textsc{Takahashi, H.},
  \textsc{Ohta, T.}, \textsc{Kawai, A.}, \textsc{Seki, K.}, \textsc{Yoshida,
  T.}, \textsc{Toyama, Y.}, \textsc{Ichikawa, H.} \& \textsc{Hasegawa, T.}
  (2007).
\newblock Gene expression analysis of soft tissue sarcomas: characterization
  and reclassification of malignant fibrous histiocytoma.
\newblock \textit{Modern pathology} \textbf{20}, 749--759.

\bibitem[{Sun et~al.(2006)Sun, Hui, Su, Vortmeyer, Kotliarov, Pastorino,
  Passaniti, Menon, Walling, Bailey et~al.}]{sun2006neuronal}
\textsc{Sun, L.}, \textsc{Hui, A.-M.}, \textsc{Su, Q.}, \textsc{Vortmeyer, A.},
  \textsc{Kotliarov, Y.}, \textsc{Pastorino, S.}, \textsc{Passaniti, A.},
  \textsc{Menon, J.}, \textsc{Walling, J.}, \textsc{Bailey, R.} et~al. (2006).
\newblock Neuronal and glioma-derived stem cell factor induces angiogenesis
  within the brain.
\newblock \textit{Cancer cell} \textbf{9}, 287--300.

\bibitem[{Tibshirani et~al.(2002)Tibshirani, Hastie, Narasimhan \&
  Chu}]{tibshirani2002diagnosis}
\textsc{Tibshirani, R.}, \textsc{Hastie, T.}, \textsc{Narasimhan, B.} \&
  \textsc{Chu, G.} (2002).
\newblock Diagnosis of multiple cancer types by shrunken centroids of gene
  expression.
\newblock \textit{Proceedings of the National Academy of Sciences} \textbf{99},
  6567--6572.

\bibitem[{Witten \& Tibshirani(2011)}]{witten2011penalized}
\textsc{Witten, D.~M.} \& \textsc{Tibshirani, R.} (2011).
\newblock Penalized classification using fisher's linear discriminant.
\newblock \textit{Journal of the Royal Statistical Society: Series B
  (Statistical Methodology)} \textbf{73}, 753--772.

\end{thebibliography}

\vskip .65cm
\noindent
University of Arizona, Tucson, AZ, 85721, USA
\vskip 2pt
\noindent
E-mail: yueniu@math.arizona.edu
\vskip 2pt

\noindent
University of Arizona, Tucson, AZ, 85721, USA
\vskip 2pt
\noindent
E-mail: nhao@math.arizona.edu
\vskip 2pt

\noindent
Beijing International Center for Mathematical Research, Peking University, Beijing, China
\vskip 2pt
\noindent
E-mail: dongbin@math.pku.edu.cn
\end{document}